# Increasing Employees' Willingness to Share: Introducing Appeal Strategies for People Analytics

Valentin Zieglmeier[1][0000-0002-3770-0321], Maren Gierlich-Joas[2][0000-0002-8731-0268], and Alexander Pretschner[1][0000-0002-5573-1201]

[1] Technical University of Munich, Munich, Germany
[2] Ludwig Maximilian University of Munich, Munich, Germany
valentin.zieglmeier@tum.de

**Abstract.** Increasingly digital workplaces enable advanced people analytics (PA) that can improve work, but also implicate privacy risks for employees. These systems often depend on employees sharing their data voluntarily. Thus, to leverage the potential benefits of PA, companies have to manage employees' disclosure decision. In literature, we identify two main strategies: increase awareness or apply appeal strategies. While increased awareness may lead to more conservative data handling, appeal strategies can promote data sharing. Yet, to our knowledge, no systematic overview of appeal strategies for PA exists. Thus, we develop an initial taxonomy of strategies based on a systematic literature review and interviews with 18 experts. We describe strategies in the dimensions of *values*, *benefits*, and *incentives*. Thereby, we present concrete options to increase the appeal of PA for employees.

## 1   Introduction and Motivation

With increasingly pervasive digitalization, significantly more data are being generated on employees. These data are a driver for people analytics (PA) in organizations, an umbrella term for "a novel, quantitative, evidence-based, and data-driven approach to manage the workforce" [1, p. 1]; [2]. PA can create actionable insights by applying data science methods to information on employees or the broader workforce. They promise to improve work by, e.g., increasing teams' efficiency or objectifying decisions [3]. Advanced PA increasingly integrate artificial intelligence, which can transform them from being descriptive in nature to predicting changes and future decision points [4]. Yet, these more advanced systems become opaque to individuals [5], leading to discomfort that hinders their adoption [6]. Those supported in their decision-making may experience fear of being replaced [7]. At the same time, employees under analysis may be subject to workplace surveillance or data misusage [8], leading to privacy concerns.

At the workplace, privacy has increased relevance for individuals. Workplace settings are very distinct from consumer settings concerning the role of privacy [9]. Due to the inherent power asymmetry between managers and employees, the risk of data misusage is elevated. Increasingly automated and opaque systems exacerbate these risks further. Therefore, recent privacy legislation, notably the European General Data Protection Legislation (GDPR), limits the use of personally identifiable information, which includes workforce behavioral data used in PA, to a minimum. In cases where the data are not required for core business processes, their use is conditional on the informed consent of the data subject (opt-in). Even if data are deemed essential, though, forcing employees to share can induce mental stress, which can make companies struggle to stay competitive [10]. Therefore, individual employees' data disclosure decisions become relevant.

Privacy calculus theory suggests that both the perceived risks (or costs) and perceived benefits of the disclosure are the basis of an individual's disclosure decision [11, 12]. Most studies we found in our search seem to address perceived risks. They focus on increasing the user's awareness and data literacy regarding what happens with their data when shared, thereby informing their disclosure decision [13]. If awareness is the goal, the concept of inverse transparency [14] can give employees oversight over the use of their data to create bi-directional transparency [15, 16]. However, those cases where personal data can serve to be beneficial to the company but are not indispensable for its processes remain critical, as they require employees to opt in when respective privacy legislation applies. Measures to increase awareness may address privacy concerns and perceived risks, but not sufficiently motivate employees to disclose their data. Instead, other strategies may be necessary to encourage data sharing.

In addition to necessary and sensible protection measures, we, therefore, recognize the need to establish motivation for sharing data in the workplace as a novel strategy. To our knowledge, there are just few prior works specifically on motivating employees to share their data. Thus, we pose our research question: *What types of appeal strategies exist that motivate employees to share personal data at the workplace?*

To answer this question, we first introduce the concepts of the digital workplace, privacy concerns and corresponding handling strategies. Next, we describe our methodology for the development of a taxonomy. Our findings synthesize the results of a systematic literature review and 18 expert interviews. For each identified dimension of appeal strategies, we present examples how it can manifest. In the discussion, the insights are reflected critically. Finally, we point out limitations, implications, and an outlook for future research in the conclusion.

Our work contributes to a holistic understanding of appeal strategies for data sharing at the workplace as we present the status quo from literature and practice. We elaborate concrete appeal strategies and contextualize them. Thereby, we show how the appeal of sharing data with PA can be understood and designed.

## 2 Conceptual Background

### 2.1 Digital Workplace

To stay competitive in disrupted markets, many organizations follow the pathway of digital transformation [17]. This transformation is defined as "organizational change that is triggered and shaped by the widespread diffusion of digital technologies" [18].

It leads to a convergent change in processes and product offerings, a transformation of work, and, ultimately, a transformation of the whole organization [19]. However, companies do not only have to compete for customers at a product level, but in times of a shortage of skilled workers, they also have to win over future employees. Consequently, leadership becomes a competitive advantage in the digital workplace. In this context, "digital transformation of work" refers to smart work, the use of digital technologies to change the workplace [20].

Smart digital workplaces can be designed by considering four patterns: digital workplace technology, workforce, new ways of working, and leadership [20]. Digital workplace technologies have emerged within the last decades and they enhance collaboration, communication, and decision-making at the workplace [19, 20]. Especially PA, which analyze workforce behavioral data to aid and automate decisions, can free employees from repetitive work [6] or support them in complex tasks to increase their efficacy [7]. Thereby, the have the potential to significantly transform the workplace.

Next to the technology-induced change, the workforce is also changing in terms of characteristics, qualifications, competencies, and mindset of managers and employees [20]. "Born-digital" millennials are familiar with many digital workplace technologies and have different expectations towards their work [20]. Thus, ways of working are adapting, employees interact differently with novel applications, and organizations work in increasingly flexible ways [20]. Virtual teams increase complexity, as they are distributed across different continents, and become the new normal, especially during times of a pandemic [21]. Leadership at digital workplaces is democratized and shared in teams [22]. Thus, technologies such as PA and novel organizational needs shape digital workplaces as well as individuals' work routines.

## 2.2    Privacy Concerns and Handling Strategies

Overall, the change in the nature of work holds benefits for employees and managers as workplaces become increasingly flexible and collaborative. However, digital workplaces also pose challenges to organizations as they entail certain risks [23]. Due to the increasing availability of employee data, new ways of control are enacted [24]. Strict control mechanisms can lead to employee stress, productivity losses, and privacy concerns. This discomfort is one of the motivators of privacy legislation, which addresses privacy concerns by limiting data processing.

With privacy concerns, we refer to individuals' belief of what happens with their data once they are disclosed [12]. These concerns are a major risk that needs to be avoided or dealt with. Generally, information privacy refers to an individual's wish to control their personal data and influence the dissemination of the data [11]. Privacy concerns arise due to the collection, processing, distribution, and usage of personal information [12, 25]. Recently, as digital workplaces have become the new normal, organizational information privacy has received the attention of many scholars [26, 27].

When considering appropriate measures to handle privacy concerns, the context has to be considered, as privacy is context-dependent [25]. We distinguish between the workplace and the consumer context, as employees experience different privacy concerns compared to consumers [9]. Moreover, as deliberated above, we differentiate the applied "awareness" or "appeal" strategy. For the awareness strategy, the goal is to increase individuals' data literacy [13]. Then, individuals are better informed about the release and use of their data, which likely leads to more restrictive handling of data. Alternatively, by increasing the appeal of sharing data, we can influence individuals' decisions towards being more inclined to share [12].

Privacy research dedicated to workplace contexts is rare; however, it is on the rise [10, 26]. Especially appeal strategies are barely applied, as we highlight in the literature review. Thus, the focus of this study is to uncover and categorize appeal strategies that motivate employees to share personal data, realizing the potential of PA.

Note that, beyond a personal benefit gained from sharing, employees may also be motivated intrinsically to contribute to the success of their company, without benefiting directly from it. For this work though, we focus on the individual benefit perspective.

## 3 Method

### 3.1 Literature Review

We start with a literature review to uncover the state of research. Our goal is to identify factors that can motivate employees to share personal data and categorize them. As the topic lies at the intersection between two disciplines, our search covers the fields of information systems (IS) and computer science (CS). We utilized Scopus for the search. Our search term was compiled to include a) terms relating to our topic of individual's data disclosure, b) terms denoting the workplace context, and c) synonyms relating to the "appeal" strategy. This resulted in the term "(disclosure decision OR sharing decision OR privacy calculus) AND (workplace OR workforce OR employee OR employer) AND (appeal OR advantage OR benefit OR incentive OR reward OR value)", covering title, abstract, and keywords.

In total, just 14 articles were found. To include additionally relevant articles, forward and backward snowballing as well as exploratory searches were employed, which added 8 articles. Furthermore, 8 articles were added manually. This resulted in 30 total articles that were analyzed based on their title and abstract. Regarding the filtering criteria, we excluded papers that were not workplace-specific, did not focus on individual data sharing, or did not follow the appeal strategy. This led to 18 articles that were re-read and discussed within the researchers' team. Applying the same filtering criteria, we arrived at 12 relevant articles for the later analysis.

### 3.2 Expert Interviews

In literature, we find mostly conceptual ideas on how to increase employees' willingness to share their data at the workplace. To enhance those results with insights from practice, we make use of semi-structured qualitative expert interviews. These allow us to observe real-life solutions from different stakeholders in a rigorous yet flexible way [28]. The interview guideline was developed and discussed within the researchers' team. It consisted of three building blocks with open-ended questions. In the introduction, we covered general questions around the interviewee's position, their use of PA, and related chances and risks. In the main part, we focused on applied appeal strategies. Finally, we gave the interviewee the opportunity for additional remarks. The interview guideline was pre-tested with one industry contact.

The interview partners were recruited through LinkedIn. For the identification of suitable interviewees, we applied the following sampling criteria: we selected experts that either a) develop PA applications, b) use PA applications, or c) consult companies in the use of PA. Users and consultants should obtain a leading function and have at least three years of experience in the field. For developers, only one year of work experience was required. The focus was on companies with a headquarter in Germany, as privacy plays an important role in Europe and privacy legislation limits data processing. Thus, we aimed to interview experts with comparable legal framing conditions. 18 individuals agreed to take part in an interview (see Table 1), representing 16 different companies from various industries, including insurance, automotive, and enterprise software. The interviews were conducted in September and October 2021 via videoconferencing systems. The interviews lasted an average of 29 minutes, excluding personal chats before or after the interview. Established guidelines were taken into consideration to avoid any biases from the interviews [28].

**Table 1.** Overview of the interview partners.

| ID | Group | Position | Experience | Recording duration |
|---|---|---|---|---|
| D1 | Developer | Head of People Analytics | 4 years | 20 min |
| D2 | Developer | Full-stack Developer | 1 year | 23 min |
| CD1 | Consultant & Developer | Consultant & Data Analyst | 3 years | 40 min |
| CD2 | Consultant & Developer | CEO | 25 years | 58 min |
| C1 | Consultant | CEO | 9 years | 28 min |
| C2 | Consultant | CEO | 6 years | 33 min |
| C3 | Consultant | Partner | 3 years | 38 min |
| C4 | Consultant | Consultant Employee Experience | 3 years | 25 min |
| C5 | Consultant | Senior Consultant | 5 years | 22 min |
| U1 | User | Personnel Controlling | 10 years | 25 min |
| U2 | User | Manager People Analytics | 4 years | 34 min |
| U3 | User | Director People Analytics | 9 years | 34 min |
| U4 | User | Manager People Development | 9 years | 25 min |
| U5 | User | Vice President HR IT Strategy | 5 years | 26 min |
| U6 | User | Head of People Analytics | 7 years | 26 min |
| U7 | User | Manager HR Reporting | 9 years | 26 min |
| U8 | User | Senior Manager Workforce | 9 years | 21 min |
| U9 | User | Head of People & Organization | 6 years | 26 min |

All interviews were recorded, transcribed verbatim, and anonymized. For the analysis, we used the tool ATLAS.ti. The coding scheme was developed iteratively and discussed within the researchers' team. For the coding categories, we built on the findings from the literature review and adapted if necessary. The quotes below were translated from German to English. An example of the coding scheme can be found in Table 2.

**Table 2.** Coding scheme.

| Dimension | Code | Example quote |
| --- | --- | --- |
| Values | Engagement | "Employees are given a chance to actually change something by themselves" (D1) |
| Benefits | Time savings | "After all, we save a lot of time and money if we implement these systems." (U7) |
| Incentives | Monetary rewards | "It's hard to set a real incentive, you can say, you'll get a 50 € Amazon voucher if you fill your skills profile, but of course that doesn't work" (U5) |

## 4 Findings

When looking at the results from our literature review and expert interviews, we can inductively derive three dimensions of appeal strategies: *values*, *benefits*, and *incentives*. Each dimension can be mapped to either the design or the usage phase of a tool (see Fig. 1). While the CS discipline mainly focuses on the design phase, the IS discipline rather addresses the usage phase and partly the design phase. Notably, incentives are the only option available in the usage phase to steer employees' behavior. They remain the only choice if the tool's design cannot be influenced.

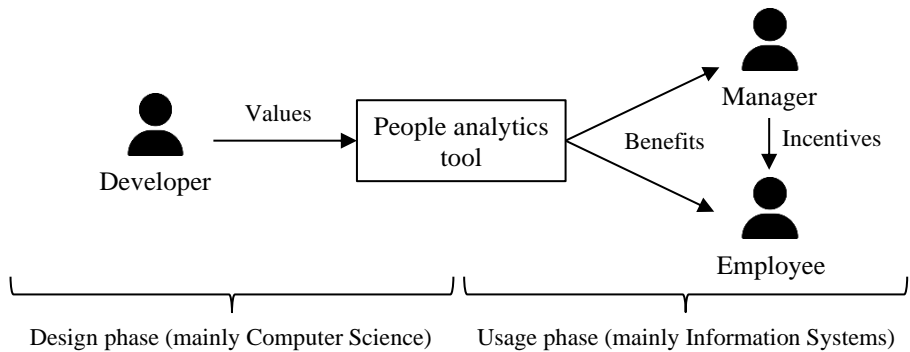

**Fig. 1.** Interplay between values, benefits, and incentives in the design phase and usage phase.

In the following, we detail the concrete definition of each dimension and describe related examples by either citing from literature or highlighting relevant interview statements. Thereby, we also reference our primary source for each.

**Values.** Some works consider values as a driver for appeal in their design. These manifest in the design of the utilized tool. In value-based engineering, generic values that users can relate to serve as the groundwork when designing software [29]. Values are inscribed into the IT artifact and, thus, clearly link the requirements of the social system and their realization in the technical system [30]. However, values are only abstract meta-requirements for the design and need to be broken down into specific design features. We found that providers of PA aim to "include certain ethical values in the systems" (C3). For most providers, it is the first step when designing a system: "For us, it is important to consider before each use case whether it is in line with our ethical principles and then go through the review process" (U2).

We identify eight categories of values which we outline in the following. First, *trust and autonomy* are central to most providers: "We don't want any negative outcomes for our employees, we want the systems to contribute to positive, employee-friendly outcomes" (U7). Systems designed to trust the users and provide them autonomy were found to be received positively in prior studies [31]. Thus, by inscribing this value, users are empowered to use PA in their own ways.

Spiekermann and Winkler [29] outline three important values to increase appeal: First, *engagement* and *psychological ownership*, meaning allowing users to directly engage with a system and develop a sense of ownership. Second, *technical market design*, meaning if the data are stored to elicit a sense of scarcity and if users can freely move them. Third, perceived *market morality*, such as if illegitimate behavior is sanctioned, which increases perceived morality and, by extension, data security [29]. The value of *engagement* was also mentioned by a developer as "employees should be given a chance to actually change something by themselves" (D1).

In contrast to personal ownership, the perceived *organizational ownership* of collected data can negatively influence users' sharing decision [32]. Conversely, data sharing increases when the more personal value of *social cohesion*, specifically *reciprocity* of sharing and support, is integrated into the system [32]. Experts confirm that some PA build on reciprocity and social cohesion as they "develop some kind of group pressure" by showing who shared the data and who did not (C1). As our interviews highlight, the reciprocity at peer level seems to be more effective in increasing employees' willingness to share compared to the hierarchical type of organizational ownership.

Still, the concrete implementation of values in the design phase of PA is challenging, and we identified only few examples. Overall, the importance of value-based engineering is confirmed by practitioners, but the specific integration of ethical values in the design phase remains somewhat vague.

**Benefits.** The largest share of analyzed literature focuses on benefits. These are inherent to the usage of the tool and irrevocably connected to it. In contrast to design-related values, benefits are realized in the usage phase. The dimension of benefits comes with the notion of providing the best possible fit between users' needs and technologies' affordances [30]. As potential risks and effort can be outweighed by the benefits of usage in privacy calculus decisions [12], providers "try to only collect data and create effort where the employee will ideally benefit from it afterward" (U5).

A benefit that is frequently highlighted in research, especially in the consumer context, is *personalization*. For example, the employer's management style can be personalized for individual employee's needs [33]. The interviews shed more light on the option of personalization. Company representatives state that "[they] do that analysis for the employees' own development" (C1) or "provide employee-tailored career paths based on the analysis" (C4).

Closely related to personalization, Cichy, Salge and Rajiv [34] describe *feedback on performance* as an immediate benefit, which they argue can reduce the need for monetary incentives. Consequently, users find it helpful to "have an overview on KPIs and to be able to compare oneself to other teammates" (U6). "Feedback is the most simple and effective, yet neglected recipe for organizations" (CD2). PA automate feedback processes as employees receive an assessment of their performance (U6). Moreover, the *social status* can be a benefit, too, as "employees can receive recognition, new job titles and prestige when using PA" (C3).

Next to these personal benefits, we find some very practical and technology-oriented benefits. PA provide *time savings* [35] as well as possible *automation of workflows* [36]. They bring "causality in the reduction to the objects of business management" (CD2). For example, if a tool stores user behavior, it could proactively suggest typical actions to save them time. Practitioners agree that PA serve as a "single point of truth" (U6) and thereby "make the life of managers and employees easier" (C2). Once implemented, PA help to "save money and time when dealing with data that are already there" (U4). Another motivating factor can be *information quality*, as Mettler and Winter have described [32]. All of these examples could be summarized as improving the work efficiency of the employee. Furthermore, Princi and Krämer [37] have outlined how *improved functionality* in the form of an increased rescue value of a smart monitoring system could represent a benefit.

Finally, an important benefit of PA lies in their potential to *initiate change*, which means that, based on shared data, change processes can be driven bottom-up: "If employees do not see in the medium and long term that something is happening based on these results, then this will also be reflected in the participation rates" (U2), whereas if "perceived relevance is high, it is one of the most important drivers" (U9). Thus, it is crucial to say "we have heard you and we base our decisions on this employee assessment" (C5). This also drives hedonic motivation when using the tool (D2).

In comparison to values, benefits are more tangible and relate to system features. In the logic of task-technology-fit, such benefits are essential to provide a good user experience, which can help overcome perceived risks during the usage [38].

**Incentives.** Third, incentives are external to the utilized tool and may be applied independently in the usage phase. Through managerial intervention, employees can be convinced to share data independent of the values and benefits of the tool. A typical example are monetary rewards [39]. Even if incentives cannot reverse existing privacy concerns that emerge from certain design characteristics, they can make the usage in the post-design phase more attractive.

Though we consulted many managers in charge of implementing PA, the majority of the interviewees struggled to think about possible incentives to facilitate data sharing. Some ideas touched upon *monetary rewards,* but interviewees "find it difficult to think about anything else, though there might be more [incentives]" (U1). The same is true for the literature, which almost exclusively discusses the characteristic of incentivizing by providing direct compensation or monetary rewards [34, 39-41]. Experts agree that monetary compensations, which are very popular in the consumer context, are perceived as problematical at the workplace: "It's hard to set a real incentive, you can say, you'll get a 50 € Amazon voucher if you fill your skills profile, but of course that doesn't work" (U5). Monetary incentives could actually worsen the situation as "the tool might be perceived even more negatively" (U5). So direct monetary rewards may need to be carefully managed in the workplace.

Alternatively, employees could be incentivized by providing explanations or *mental support by their manager* [42]. If managers act as role models, saying "I also enter my data, there is nothing bad about it" (U4), employees are more open towards sharing their data. One could imagine a skill management system that is used by managers to specifically support employees that rank lower. Next, *gamification* models, such as bonus points that can be collected, are assessed positively by the interviewees (C3). However, gamification remained a hypothetical incentive that has not yet been broadly applied. Lastly, managers handle PA by incentivizing employees to share their data in return for individual *training & coaching* opportunities. With these, the idea is to "give the individual something back for themselves, but not in the form of money" (U8). In one case, shared employee data were used for a so-called "potential conference" where future organizational leaders were promoted based on these insights. Without revealing their data, employees could not be considered in the promotion (U6).

In conclusion, the measures in the usage phase in terms of incentives are scarcely covered. Our empirical findings highlight that current efforts on increasing employees' willingness to share are mostly expressed in the design phase but the daily interaction with the tools has not yet been questioned.

**Taxonomy.** The resulting taxonomy below (see Table 3) is a synthesis of the findings from literature and interviews. It categorizes all above-described examples in the dimensions of *values*, *benefits*, and *incentives*.

Overall, it is our aim to provide an initial taxonomy for the field of handling employee privacy concerns at the workplace with appeal strategies, as, to the best of our knowledge, no prior works exist. Therefore, the taxonomy is meant to be a stepping stone for future research and needs to be evaluated further.

We created it by first compiling the results from literature and building an initial taxonomy. Then, we supplemented it with the results from the interviews. Mostly, the findings from the interviews overlapped with those from literature. Three examples were added after the interviews, namely the benefit *change initiation*, as well as the incentives *gamification* and *training & coaching*. To assign them to a dimension, we deductively matched them based on our underlying model.

**Table 3.** Taxonomy of appeal strategies for people analytics.

| Dimension | Examples | | | |
|---|---|---|---|---|
| *Values* | Trust | Autonomy | Engagement | Psychological ownership |
| | Market morality | Technical market design | Organizational ownership | Social cohesion & reciprocity |
| *Benefits* | Personalization | Feedback on performance | Social status | Time savings |
| | Automation | Information quality | Improved functionality | Change initiation |
| *Incentives* | Compensation & monetary rewards | Mental support by manager | Gamification | Training & coaching |

## 5  Discussion

PA systems at the workplace are on the rise. They will change decision making and individuals' work routines. Therefore, they need to be managed carefully to avoid potential downsides while increasing employees' willingness to share. Two main strategies emerge: awareness and appeal. To ensure informed consent, true awareness of individuals is required, and working towards it seems to be a sensible first step. Existing literature as well as the results from our interviews show a clear focus on awareness strategies. One potential explanation is that many employees are skeptical about PA, as they lack digital capabilities and a data mindset (C2). Therefore, managers first need to reduce fear by increasing awareness about the collected data and their purpose, as interview partner D1 states: "No, I don't want to track you, I am more interested in finding patterns and optimizing workflows – that's what I am advertising a lot" (D1). At the same time, though, awareness alone may not be enough to motivate data sharing in the end, due to simple inertia or a lack of positive motivation to share. Therefore, appeal strategies become relevant. Yet, we need to consider the various points of contention that may arise.

First, we should not overlook the ethical issues with this situation. Roßnagel, Pfitzmann and Garstka [43] assume that due to the inherent dependency, true consent is impossible at the workplace. Therefore, it remains a question whether employee privacy concerns may simply need to be accepted in some cases, such as when handling sensitive data. This also depends on the cultural context that is being investigated, as legal conditions and the perception of privacy risks vary significantly. The expectation to encode values in software, for example, is partially enshrined in European privacy legislation. In the end, the perceived risks depend substantially on the advantages that the data usage has for individuals. Therefore, we can consider the work on positive motivation for data sharing to be relevant.

Considering appeal strategies, we find that a focus lies on the inherent benefits the tools provide. Their connection to the provided data is often intuitive, and developers are naturally motivated to focus on the benefits their tools have. Contrary to that, values and incentives need to be considered explicitly. In literature, we find two main research streams that cover these aspects. Spiekermann and Winkler describe value-based engineering as a practice to ensure that systems are not merely useful (provide benefits), but "support what is good, true, beautiful, peaceful and worthy in life" [29, p. 1]. However, it is up to the designer to define desirable values that shape the tools, which could lead to unintended biases. Moreover, as encoded values are rather abstract, it is important to consider the effect of implemented values in practice, as they may not even be actively noticed in typical use. Incentives, on the other hand, are more directly experienced. They can generally be regarded as a managerial lever, applied after a tool has been deployed. For more intricate incentive mechanisms and automated distribution, incentive engineering could be a relevant field of research [44]. Here, mechanisms are embedded into tools to provide incentives when, e.g., data are shared. A relevant point of discussion then is if this kind of automated system becomes compelling to be gamed, shifting motivation to achieving high scores instead of contributing value. Conversely, if the system does not detect that an incentive should be distributed, this could demotivate further sharing. Furthermore, advanced PA are perceived as biased and not always transparent [2]. Giving them the power to incentivize employees can also be seen as a risk. Therefore, we should acknowledge the effect of such automation on the managerial agency. If managers are still meant to hold responsibility, they should be enabled to adapt the incentive mechanisms or have final say in the distribution.

## 6 Conclusion

### 6.1 Summary, Theoretical Contribution, and Practical Implications

The goal of this work is to identify appeal strategies that enable PA at the workplace, which differ from established awareness strategies. Guided by the research question "What types of appeal strategies exist that motivate employees to share personal data at the workplace?", we choose the approach to develop a taxonomy. The taxonomy enables us to display the strategies in a structured way as we derive the dimensions of values, benefits, and incentives, including different characteristics. Value-centric design of PA is an emerging trend which needs specification and concrete features. PA provide numerous benefits in use which need to be communicated via success stories. The managerial levers in the usage phase are still vague and incentives from the consumer context cannot be simply applied to the workplace. While values are aiming at the design phase, thus touching upon the field of CS, incentives are unfolding in the usage phase and are mainly investigated from IS perspective.

Our study holds the following contributions for academia: First, with this multidisciplinary work, we bridge the disciplines of IS and CS and are able to investigate the design phase and the usage phase of digital workplace technologies equally. Second, our focus on the workplace setting differs from most prior studies. As the conditions at the workplace are very distinct from the consumer context, established strategies to handle privacy concerns cannot simply be transferred. With our analysis, we take these specific requirements into account and contribute to the literature on privacy at the workplace. Third, our approach to tackle data sharing at the workplace is novel, as the focus in the past has mostly been on awareness strategies. With our work, we underline the need to apply appeal strategies.

We believe that practitioners can benefit from our study as well. In digital times, employee privacy concerns are rising and threaten productive collaboration and employee's mental health. Hence, managing these concerns is a crucial managerial challenge. For users of PA, we identify concrete managerial levers that can be applied in the usage phase to increase perceived appeal. If this overarching goal is achieved, employees are more willing to share their data, and organizations can benefit from novel use cases of the data. For developers of PA, we stress the importance of value-based engineering and benefit-driven design. If the proposed strategies are implemented as a foundation in their tools, we believe this can reduce employee privacy concerns and increase their appeal. Thus, from a more general perspective, our study assists in designing the interaction between humans and IS at the workplace.

### 6.2 Limitations and Outlook

Despite being rigorously conducted, our approach is methodologically limited. First, the literature review only covers works that are matched by the terms we chose. To counteract this issue, we employed forward and backward snowballing. Second, the interview partners have been subject to biases. Therefore, we took care to choose a diverse set of perspectives. Moreover, for the interviews, we explicitly limited our focus on the European market, specifically Germany. This was a conscious choice to shed light on a single cultural and legal context, but it means that our results may not be representative of other countries and cultures. Finally, the taxonomy is an initial version and has not yet been applied to broader company cases.

Considering these limitations, various next steps seem promising. First, we suggest validating the taxonomy further by using it in distinct cases and different cultural settings. Focus group discussions including employees affected by the appeal strategies could help in the evaluation of the taxonomy. Furthermore, we regard the idea of explicitly embedding values or incentives into tools as an interesting follow-up. There are still open questions whether this would influence individuals' data sharing decisions, and future research could consider the implications for executive agency and actions. Hence, we consider the work a stepping stone for future design-oriented studies in the fields of value-based engineering or incentive engineering that can facilitate the management of PA at the workplace.

**Acknowledgments.** This work was supported by the German Federal Ministry of Education and Research (BMBF) under grant no. 5091121.